# Synthesis and Magnetic Properties of Cobalt Ferrite ($CoFe_2O_4$) Nanoparticles Prepared by Wet Chemical Route

K. Maaz, Arif Mumtaz[+], S.K. Hasanain, Abdullah Ceylan[*]

*Department of Physics, Quaid-i-Azam University, Islamabad, Pakistan*
*[*]Department of Physics and Astronomy, University of Delaware, Newark*

## Abstract

Magnetic nanoparticles of cobalt ferrite have been synthesized by wet chemical method using stable ferric and cobalt salts with oleic acid as the surfactant. X-ray Diffraction (XRD) and Transmission Electron Microscope (TEM) confirmed the formation of single phase cobalt ferrite nanoparticles in the range 15-48nm depending on the annealing temperature and time. The size of the particles increases with annealing temperature and time while the coercivity goes through a maximum, peaking at around 28nm. A very large coercivity (10.5kOe) is observed on cooling down to 77K while typical blocking effects are observed below about 260K. The high field moment is observed to be small for smaller particles and approaches the bulk value for large particles.

*PACS: 75.50Gg; 75.50Tt; 75.70Rf*

*Keywords: Ferrite Nanoparticles; Surface anisotropy; High coercivity magnetic materials;*

[+]Correspondence Author: arif@qau.edu.pk (A. Mumtaz)

# Introduction

Recently metal-oxide nanoparticles have been the subject of much interest because of their unusual optical, electronic and magnetic properties, which often differ from the bulk. Cobalt ferrite ($CoFe_2O_4$) is a well-known hard magnetic material with high coercivity and moderate magnetization. These properties, along with their great physical and chemical stability, make $CoFe_2O_4$ nanoparticles suitable for magnetic recording applications such as audio and videotape and high-density digital recording disks etc. [1, 2]. The magnetic character of the particles used for recording media depends crucially on the size, shape and purity of these nanoparticles. These particles should be single domain, of pure phase, having high coercivity and medium magnetization. Hence the need for developing fabrication processes that are relatively simple and yield controlled particle sizes.

Conventional techniques for preparation of nanoparticles include sol-gel processing, hot spraying, evaporation condensation, matrix isolation, laser-induced vapor phase reactions and aerosols. Generally, in most types of nano-particles prepared by these methods, control of size and size distribution is not possible [1]. In order to overcome these difficulties, nanometer size reactors for the formation of homogeneous nanoparticles of cobalt ferrite are used. To protect the oxidation of these nanoparticles from the atmospheric oxygen and also to stop their agglomeration, the particles are usually coated and dispersed in some medium like sodium dodecyl sulfate (NaDS) or oleic acid [3, 4].

In general, the preparation methods for $CoFe_2O_4$ nanoparticles have been quite involved requiring special techniques to prevent agglomeration [5] or microwave assisted synthesis [6]. In this paper, we have presented the synthesis of cobalt ferrite ($CoFe_2O_4$) nanoparticles by wet chemical method (coprecipitation) along with heat treatment at 600°C. The size and size distribution was controlled by controlling the nucleation and growth rates. Smaller and uniformly disturbed particles were obtained if the nucleation rate was higher than the growth rate. Large pH values in the range (11-14) were used in accordance with predictions where high production yields are expected for large pH values. The advantage of this method over the others is that the control of production of ferrite particles, its size and size distribution is relatively easy and there is no need of extra mechanical or microwave heat treatments.

The size and size distribution of the particles prepared by this method was studied by XRD and TEM. The dependence of the particle size on the annealing temperature and annealing time was also studied. Finally, various magnetic properties of the particles have been studied as functions of field, temperature and size.

## Experimental procedure

**Materials**

Ferric chloride and cobalt chloride (98 +% purity) and sodium hydroxide were used. Oleic acid of HPCL grade was used as surfactant. All the materials were reagent grade and used without further purification. Double distilled, de-ionized water was used as a solvent.

**Procedure**

0.4M (25ml) solution of iron chloride and a 0.2M (25ml) of cobalt chloride solutions were mixed in double distilled, de-ionized water. Deionized distilled water was used as a solvent in order to avoid the production of impurities in the final product. 3M (25ml) solution of sodium hydroxide was prepared and slowly added to the salt solution dropwise. The pH of the solution was constantly monitored as the NaOH solution was added. The reactants were constantly stirred using a magnetic stirrer until a pH level of 11-12 was reached. A specified amount of oleic acid was added to the solution as a surfactant and coating material [1]. The liquid precipitate was then brought to a reaction temperature of $80^oC$ and stirred for one hour. The product was then cooled to room temperature. To get free particles from sodium and chlorine compounds, the precipitate was then washed twice with distilled water and then with ethanol to remove the excess surfactant from the solution. To isolate the supernatant liquid, the beaker contents were then centrifuged for fifteen minutes at 3000 rpm. The supernatant liquid was then decanted, and then centrifuged until only thick black precipitate remained. The precipitate was then dried overnight at $100^oC$. The acquired substance was then grinded into a fine powder. At this stage the product ($CoFe_2O_4$) contains some associated water (upto 10 wt%), which was then removed by heating at $600^oC$ for ten hours. The final product obtained was then confirmed by X-ray diffraction etc to be magnetic nanoparticles of cobalt ferrite ($CoFe_2O_4$) with inverse spinel structure (details below).

## Results and discussion

The X-diffraction pattern (Fig.1) of the calcined powder synthesized using this route shows that the final product is $CoFe_2O_4$ with the expected inverse spinel structure. No other phase/impurity was detected. The size of the particles was determined by Scherrer formula using first two strongest peaks. The average sizes of the particles calcined at $600^oC$ were found to be 15nm, 17.5nm, 21± 3nm with variation in sizes being introduced by controlling the rate of mixing of NaOH with the salt solution. Slow rates of mixing resulted in larger size particles as the growth rate begins to exceed the nucleation rate. By annealing 15nm particles at 800, 900, 850 and $1000^oC$ respectively for 10 hours, particle sizes of 24, 26, 32 and 38± 3nm were obtained. Finally, on further annealing the 17.5nm particles, described above, at $1000^oC$ for 6 and 10 hours respectively, 42 and 48 ± 3nm particles were obtained. Thus the size variation and control has been achieved by both the rate of reaction and the annealing conditions.

Fig. 2 shows the TEM images of $CoFe_2O_4$ nanoparticles calcined at $600^oC$ for 10 hours (with average crystallite size of about 21 nm as determined by XRD). The size distribution of these nanoparticles as observed in TEM images is shown in Fig. 3. The distribution seems to be symmetric (Gaussian) about 21 nm, with particles of sizes 16-26nm for this specimen. The maximum number lie between 20 and 22nm, peaking at 21nm, in good agreement with XRD crystallite size. Most of the particles appear spherical in shape however some elongated particles are also present as shown in the TEM images. Some moderately agglomerated particles as well as separated particles are present in the images. Agglomeration is understood to increase linearly with annealing temperature and hence some degree of agglomeration at this temperature ($600^oC$) appears unavoidable.

Fig. 4 shows the correlation between the particle size and annealing temperature. The size of the particles is observed to be increasing linearly with annealing temperature. It appears that that increase in size with temperature becomes rapid between $800-900^oC$ and appears to be slowing down above $900^oC$. While annealing generally decreases the lattice defects and strains, however it can also cause coalescence of crystallites that results in increasing the average size of the nanoparticles [7]. The inset in Fig. 4 shows the dependence of particle size on annealing time at a fixed annealing temperature of 600$^o$C. The particle size appears to increase almost linearly with annealing time, most likely due to the fact that longer annealing time enhances the coalescence process resulting in an increase in the particle size. Thus it appears that particle size may be controlled by varying either of the two parameters i.e. annealing temperature and time.

Magnetic characterization of the particles was done using vibrating sample magnetometer (VSM), between room temperature and 77K, with maximum applied field upto 15kOe (see Fig. 5 for M-H loops). For the 24nm size particles the coercivity at room temperature was 1205Oe while at 77K it had increased to ~11kOe. The saturation magnetization ($M_S$) obtained at room temperature was found to be 68emu/g and remanent magnetization ($M_r$) was 31.7emu/g, while at 77K the values for the same parameters were 40.8emu/g and 34.4emu/g respectively. The very large coercivity and low saturation magnetization at 77K are consistent with a pronounced growth of magnetic anisotropy inhibiting the alignment of the moment in an applied field. The remanence ratios at these temperatures indicate the same feature, rising from 0.47 to 0.84 at 77K. The value of remanence ratio of 0.47 is very close to that expected (0.5) for a system of nonintracting single domain particles with uniaxial anisotropy [8] even though cobalt ferrite itself has a cubic structure. The existence of an effectively uniaxial anisotropy in magnetic nanoparticles has been attributed to surface effects [9] as evidenced by simulations of nanoparticles. Surface effects also tend to lead to large anisotropies. Kodama et al [9] have argued that due to the disorder near the surface, the typical two sublattice picture for antiferromagnetic or ferrimagnetic nanoparticles appears to break down and multiple sublattice picture appears to hold. There appears to be a situation where several different spin configurations e.g. 2, 4, 6 sublattice models have very similar energies and hence multiple ground states are possible e.g. in a spinglass. The interaction between the core and surface spins leads to a variety of effects including large anisotropy and exchange bias effects. Such effects are observed in these particles and are to be reported separately.

It is common to find higher effective anisotropy values in magnetic nanoparticles as compared to their bulk counterpart. From a fit of the magnetization data at high fields using the Law of Approach to Saturation,

$$\chi = \partial M/\partial H \cong \alpha K_{eff}^2 / M_s H^3 ,$$

we obtained the value of the anisotropy constant $K_{eff}$= 3.8×10$^6$ ergs/cc, where α = 0.533 for uniaxial anisotropy [8]. This value is somewhat but not very much higher than the value of 1.8-3.0×10$^6$ ergs/cc [10] for bulk cobalt ferrite, which has a room temperature coercivity of 750-980Oe [11].

The coercivity of the nanoparticles was also studied as a function of particle size. Fig 6. shows the coercivity as a function of particle size at room temperature (300K). The Gaussian fit to the data shows the coercivity increases with size rapidly, attaining a maximum value of ~1250Oe at 28 nm and then decreases with size of the particles. This decrease at larger sizes could be attributed to either of two reasons. Firstly, it may be due to the expected crossover from single domain to multidomain behavior with increasing size. Secondly such an effect can arise from a combination of surface anisotropy and thermal energies. The former effect is expected in CoFe$_2$O$_4$ particles for a size close to 50nm [12, 13] that is significantly higher than the critical size of 28nm that we observed. The latter source of the effect therefore is considered as more likely explanation for the peak. We understand the initial increase of the coercivity with decreasing size as being due to the enhanced role of the surface and its strong anisotropy, as opposed to the weaker bulk anisotropy. This rise is followed by a decline at small enough sizes when the product of the anisotropy energy and volume becomes comparable to the thermal energy, leading to thermally assisted jumps over the anisotropy barriers. It is also likely that the two processes are operating simultaneously and the single domain effects may not be excluded, however the dominant role will be of the surface effects for smaller particles. The decrease of H$_c$ at d ≥ 40nm may very well have a contribution from the development of domain walls in the nanoparticles. These aspects shall be discussed in detail elsewhere [14]. The magnetization of different size nanoparticles is shown as a function of temperature in Fig. 7. The samples were cooled to 77K without application of external magnetic field (ZFC). Following the cooling a field of 5kOe was applied and magnetization was recorded as function of temperature up to 300K. A peak in the magnetization is evident in each case with the exact position of the peak depending on the size. It is understood that in the ZFC mode the magnetization of a collection of nanoparticles may go through a peak as the particles' moments become blocked along the anisotropy axes. The temperature at which maximum magnetization is achieved, is defined as the blocking temperature ($T_B$). This temperature is a function of applied field and typical time scale of measurement.

The inset in Fig. 7 shows the effect of size on the blocking temperature. While there is a clear increase in the blocking temperature with size, it is also apparent that this increase is very rapid in the beginning (at smaller sizes) and thereafter the increase becomes very slow appearing to reach a maximum at T~270K (for H=5kOe). The larger particles seem to be blocked at high temperatures as compared to the smaller particles at the same field. For larger particles, the larger volume causes increased anisotropy energy which decreases the probability of a jump across the anisotropy barrier and hence the blocking is shifted to a higher temperature. From the data it appears that above about 24nm the particle blocking becomes relatively insensitive to size.

Fig. 8. shows the dependence of saturation magnetization on particle size. The $M_S$ values obtained for our samples vary between 53 to 79.5emu/gm. The maximum value is 79.5emu/gm for 48nm particles close to the bulk value of 80.8emu/gm for $CoFe_2O_4$.

The saturation magnetization increases consistently with size. For small particles the value of Ms is significantly lower than the bulk value of 80emu/gm while for the size of ~48nm the magnetization has attained the bulk value. Once again we see a very sharp increase in the magnetization between the sizes of 15-28nm while there is a slower increase thereafter, as in the case of coericivity and blocking temperature. The decrease in $M_S$ at small sizes is attributed to the effects of the relatively dead or inert surface layer that has low magnetization. This surface effect becomes less significant with increasing sizes and above 48nm seems to be no longer relevant to the bulk magnetization.

## Conclusion

In this paper, we have presented the synthesis of $CoFe_2O_4$ nanoparticles in the range 15-48nm. The size of the nanoparticles were measured both by XRD and TEM and were in very good agreement with each other indicating that there was no agglomeration and that the size distribution of the prepared nanoparticles was small. The size of the nanoparticles appeared to increase linearly with annealing temperature and time most probably due to coalescence that increases with increasing temperature of anneal. It is evident that particle size and its distribution may be controlled both by controlling the rate of reaction and the annealing temperature and time period. The very large coercivity and low saturation magnetization at 77K in comparison with room temperature appear to be due to a pronounced growth of magnetic anisotropy at low temperatures. The observed magnetization remanence ratio of 0.47 at room temperature (very close to the value of 0.5 typical of a system of noninteracting single domain particles) suggests that $CoFe_2O_4$ nanoparticles exhibit an effective anisotropy that is uniaxial. The effective uniaxial anisotropy in magnetic nanoparticles has been explained as arising from surface effects that also lead to large anisotropy energy in nanoparticles. The coercivity shows a peak with particle size at a value much smaller than the single domain limit and is attributed to the onset of thermal effects at small enough particle sizes. We find that for smaller particles the saturation magnetization had a value that was significantly lower than the bulk value while the larger size particles have values approaching those of the bulk. The smaller value of $M_S$ in smaller particles is attributed to the greater fraction of surface spins in these particles that tend to be in a canted or spin glass like state with a smaller net moment.

## Acknowledgment

K.M and A. M wish to acknowledge the Higher Education Commission (HEC) for providing PhD fellowship and Research Grant (*No. 20-74/Acad(R)/03*) for enabling this work. S.K.H also acknowledges the H.E.C for Research Grant (*No. 20-80/Acad(R)/03*) on ferromagnetic nanomaterials.

## References

[1] V. Pallai and D.O. Shah, Journal of Magnetism and Magnetic Materials **163,** 243-248, (1996)

[2] R Skomski, J. Phys.: Condens. Matter **15**, R1-R56, (2003).

[3] T. Feried, G. Shemer, G. Markovich, Adv. Mater. **13**, No.15, pp. 1158-1161, August 3, 2001.

[4] C. Liu, A. J. Rondinone, Z. J. Zhang, Pure Appl. Chem., Vol. **72**, Nos. 1-2, pp. 37-45, (2000).

[5] H. Yang, Xiangchao Zhang, Aidong Tang, Guanzhou Qiu, Chemistry Letters Vol. **33**, pp. 826-827, No. 7 (2004).

[6] R Honrada, R. Seshadri, A. Risbud, NNUN REU Program at Nanotech, Nanofabrication Facility at UC, Santa Barbara, pp. 106-107, (2003).

[7] T. P. Raming, A. J. A. Winnubst, C. M. Van Kats, P. Philipse, Journal of Colloid and Interface Science **249**, 346-350, (2002).

[8] Takahiro Ibusuki, Seijyu Kojima, Osamu Kitakami, Yutaka Shimada, IEEE Transactions on Mangnetics Vol. 37, pp. 2223-2225, No. 4, July, 2001.

[9] R. H. Kodama, Journal of Magnetism and Magnetic Materials **200**, pp 359-372, (1999).

[10] L. D. Tung, V. Kolesnichenko, D. Caruntu, N. H. Chou, C. J. O'Connor, and L. Spinu, Journal of Applied Physics, Volume **93**, Number 10, May, 2003.

[11] D.J. Craik, "Magnetic Oxides", Part 2, pp. 703, John Wiley & Sons, London, (1975).

[12] W. J. Schuele, Y. D. Deet Screek, W.W. Kuhn, H. Lamprey, C. Scheer (Eds) Ultrafine Particles Wiley (Newyork) p.218, (1963).

[13] A.E. Berkowitz, W.J. Schuele, J. Appl. Phys. **30**, 1345(1959): C.N. Chinnasamy, B. Jeyadevan, K. Shinoda, K. Tohji, D. J. Djayaprawira, M. Takahashi, R. Justin Joseyphus, A. Narayanasamy, Appl. Phys. Lett. **83** 2862 (2003).

[14] A. Mumtaz, K. Maaz, B. Janjua, S.K. Hasanain, J. Phys. Cond. Matt. (submitted for publication, 2006).

# List of Figures

Fig. 1: X-ray diffraction pattern (Cu $K_\alpha$-radiation) of $CoFe_2O_4$ nanoparticles prepared by wet chemical method, after calcination at $600^oC$ for 10 hrs with average crystallite size of about 21 nm.

Fig. 2: TEM micrograph of $CoFe_2O_4$ nanoparticles prepared, after calcination at $600^oC$ for 10 hrs with average crystallite size of 21 nm.

Fig. 3: Size distribution (histogram) of $CoFe_2O_4$ nanoparticles from TEM images ($600^oC$ calcination for 10 hrs)

Fig. 4. Particle size (nm) as a function of annealing temperature ($^oC$) for $CoFe_2O_4$ nanoparticles. The inset shows the correlation between particle size and annealing time, in hrs (at $T=600^oC$).

Fig. 5. Hysteresis loops for 24nm $CoFe_2O_4$ nanoparticles at room temperature (300K) and 77K at maximum applied field of 15 kOe.

Fig. 6. The correlation between the coercivity ($H_C$) and mean particle diameter (nm), at room temperature and applied field of 15kOe.

Fig: 7. Temperature dependence of magnetic susceptibility for zero-field cooled (ZFC) $CoFe_2O_4$ nanoparticles at applied field of 5 kOe. The inset is the dependence of blocking temperature ($T_B$) on particle size (nm).

Fig. 8. Saturation magnetization ($M_S$) as function of particle size (nm) at maximum applied field of 15kOe.

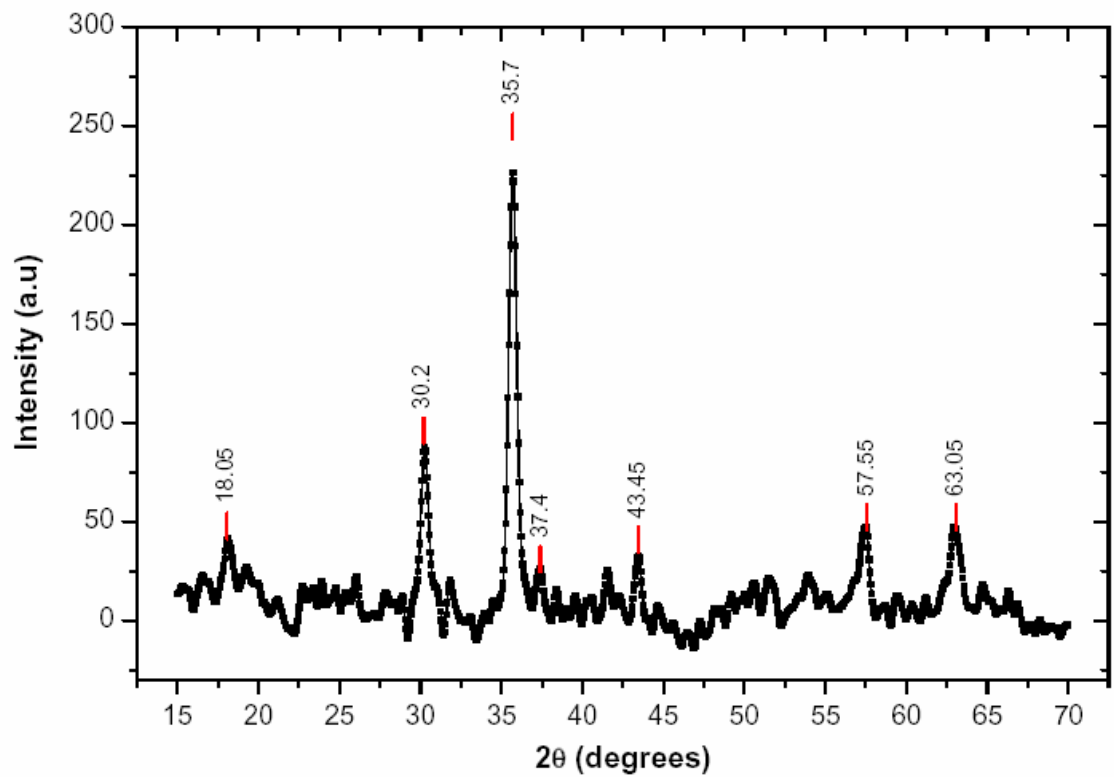

Fig. 1

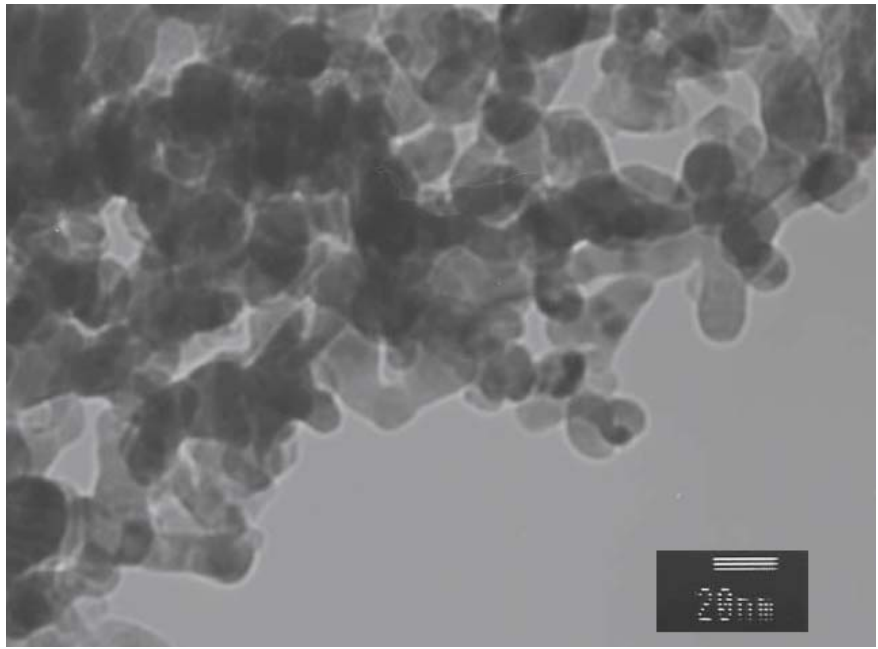

Fig. 2

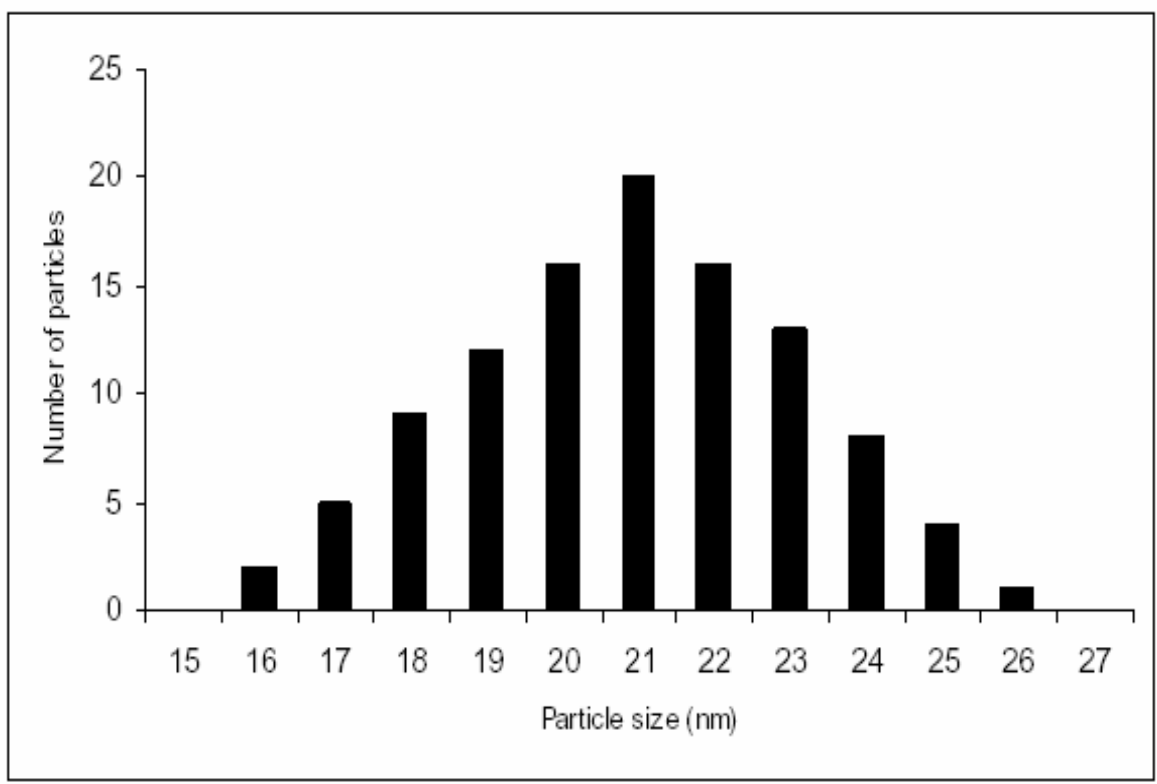

Fig. 3

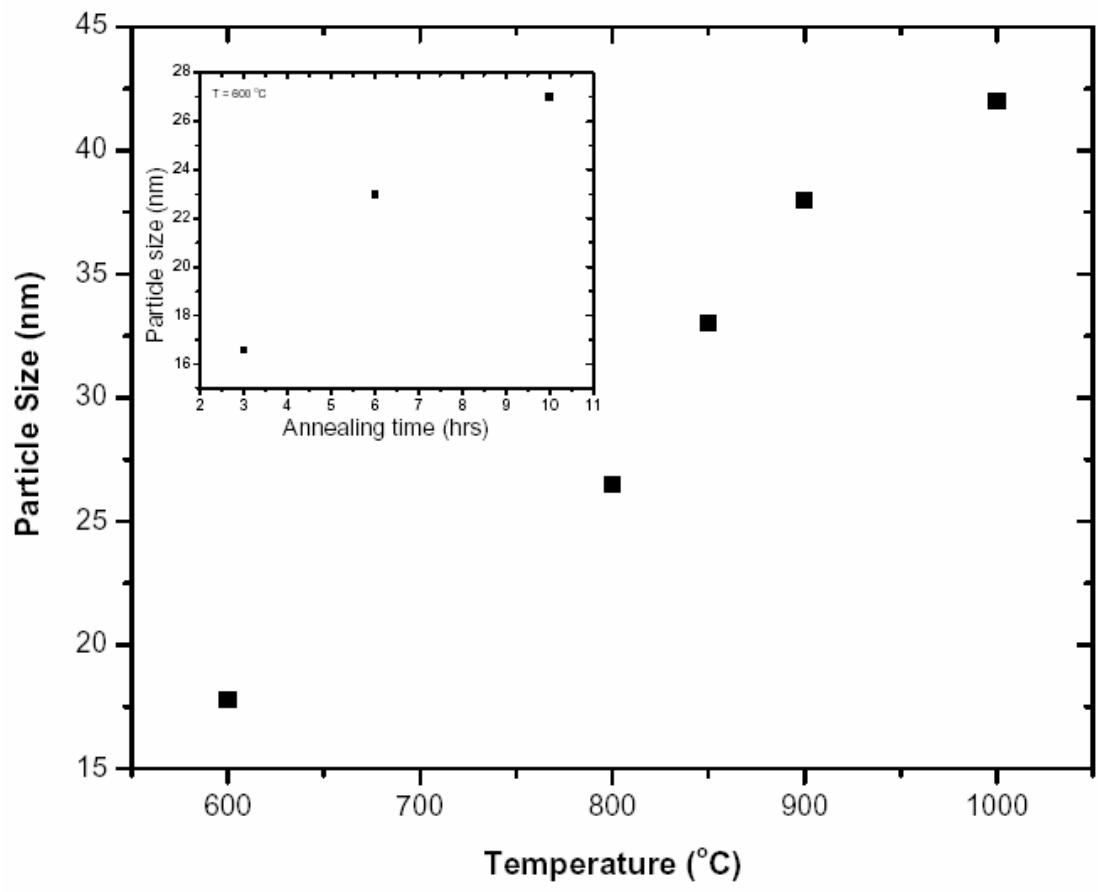

Fig. 4

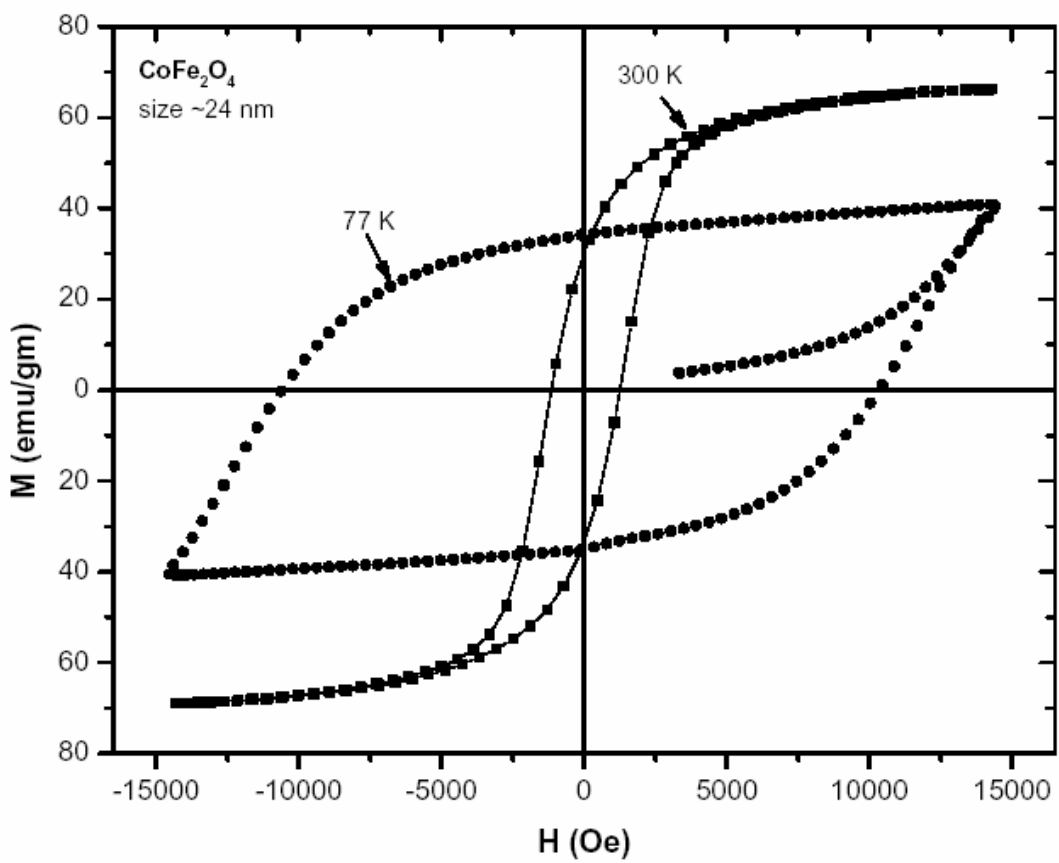

Fig. 5

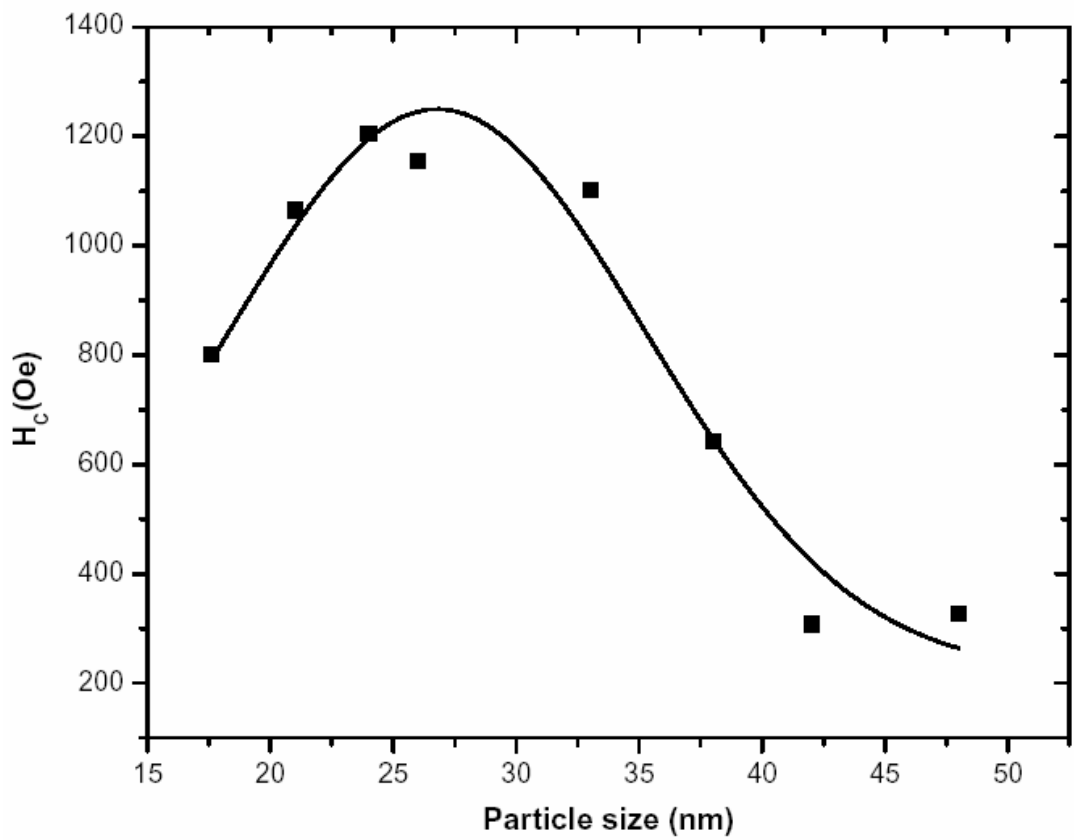

Fig. 6

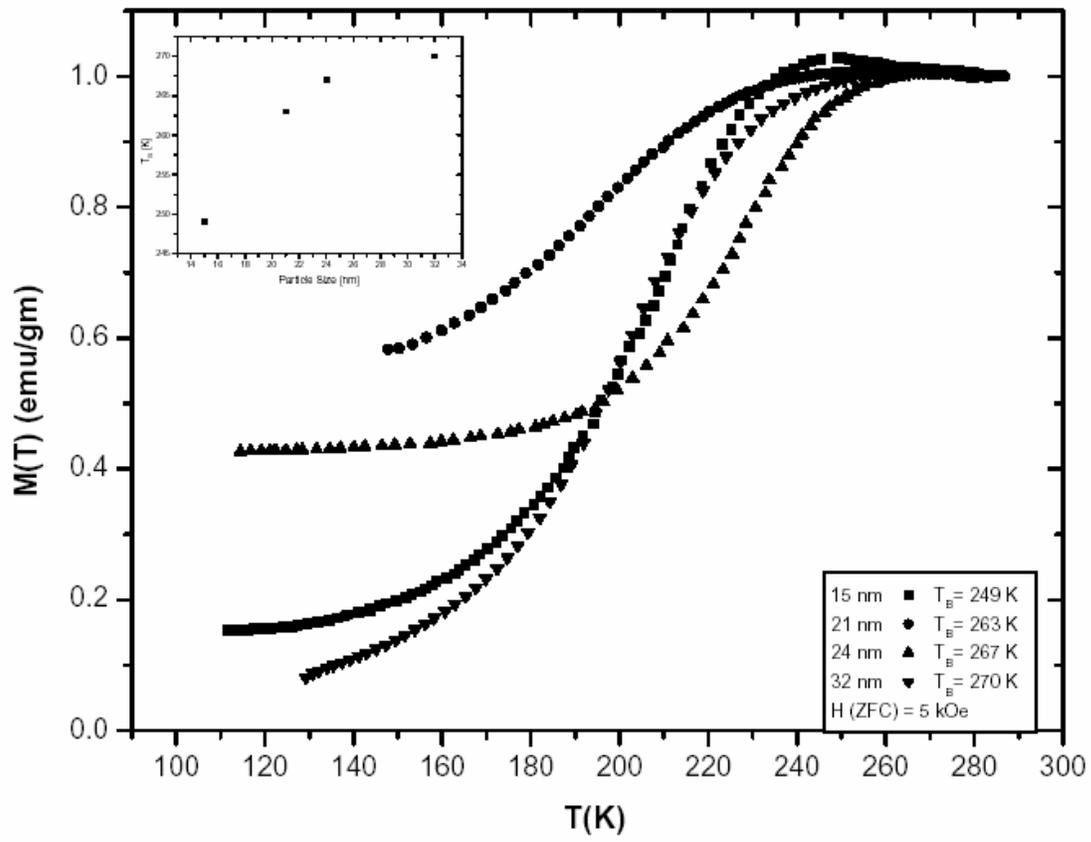

Fig. 7

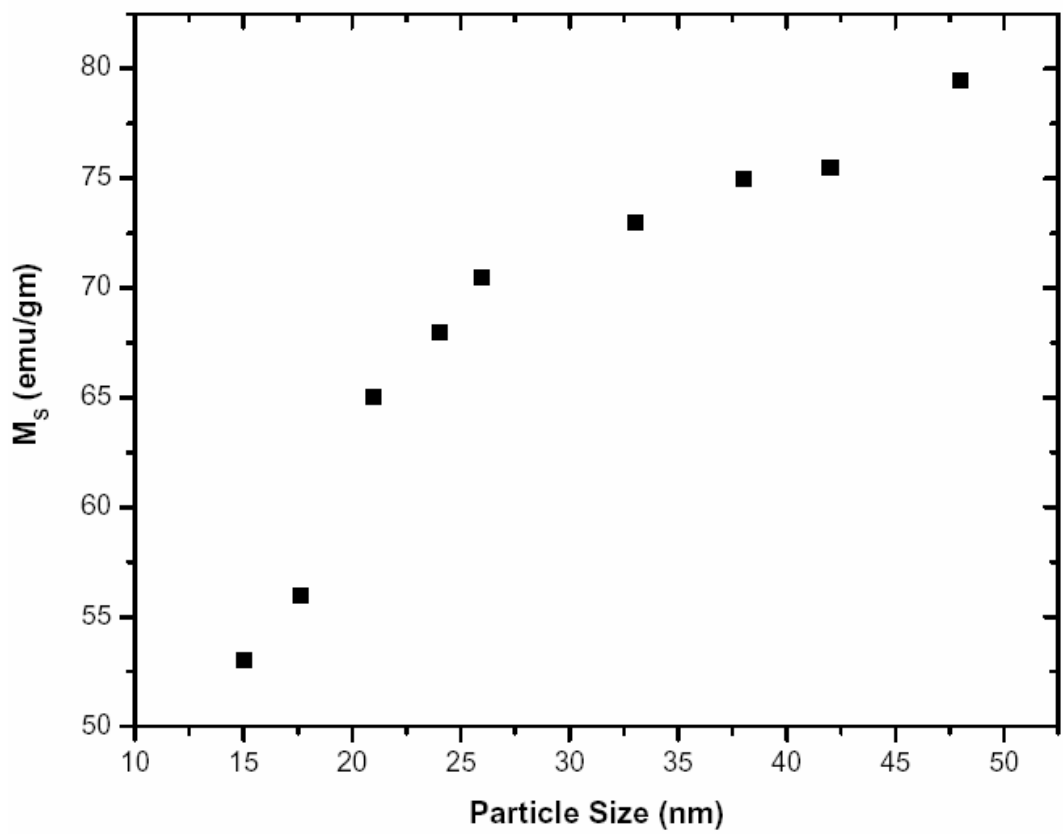

Fig. 8